\documentclass[final,5p,authoryear,twocolumn]{elsarticle}

\usepackage[british]{babel}
\usepackage{graphicx}
\usepackage{longtable}
\usepackage{amssymb}
\usepackage{color}
\usepackage{overpic}
\usepackage{hhline}

\usepackage{lineno}

\newcommand{\reftab}[1]{Table~\ref{#1}}

\newcommand{\reffig}[1]{Figure~\ref{#1}}

\newcommand{\comment}[1]{}

\newcommand{\koeroes}{K\"or\"os}

\newcommand{\kma}{\ensuremath{\mathrm{km\,a}^{-1}}}
\newcommand{\sqkma}{\ensuremath{\mathrm{km}^{2}\,\mathrm{a}^{-1}}}

\newcommand{\science}[2]{\ensuremath{#1\cdot 10^{#2}}}

\journal{Journal of Archaeological Sciences}

\begin{document}
\begin{frontmatter}

\title{A simulation of the Neolithic transition in Western Eurasia}

\author[hzg]{Carsten Lemmen\corref{lab:cor}}
\ead{carsten.lemmen@hzg.de}
\cortext[lab:cor]{Tel +49\,4152\,87-2013, Fax~-2020} 
\address[gkss]{Helmholtz-Zentrum Geesthacht, Institut f\"ur K\"ustenforschung, Max-Planck Stra\ss e~1, 21501~Geesthacht, Germany}
\author[rgzm]{Detlef Gronenborn}
\address[rgzm]{R\"omisch-Germanisches Zentralmuseum, Ernst-Ludwig Platz~2, 55116 Mainz, Germany}
\author[hzg]{Kai W. Wirtz}

\begin{abstract} 
Farming and herding were introduced to Europe from the Near East and Anatolia; there are, however, considerable arguments about the mechanisms of this transition.  Were it the people who moved and either outplaced or admixed with the indigenous hunter-gatherer groups? Or was it material and information that moved---the Neolithic Package---consisting of domesticated plants and animals and the knowledge of their use? The latter process is commonly referred to as cultural diffusion and the former as demic diffusion. Despite continuous and partly combined efforts by archaeologists, anthropologists, linguists, palaeontologists and geneticists, a final resolution of the debate has not yet been reached.
In the present contribution we interpret results from the Global Land Use and technological Evolution Simulator (GLUES). This mathematical model simulates regional sociocultural development embedded in the geoenvironmental context during the Holocene. We demonstrate that the model is able to realistically hindcast the expansion speed and the inhomogeneous space-time evolution of the transition to agropastoralism in western Eurasia. In contrast to models that do not resolve endogenous sociocultural dynamics, our model describes and explains how and why the Neolithic advanced in stages.
We uncouple the mechanisms of migration and information exchange and also of migration and the spread of agropastoralism. We find that (1)~an indigenous form of agropastoralism could well have arisen in certain Mediterranean landscapes but not in northern and central Europe, where it depended on imported technology and material; (2)~both demic diffusion by migration and cultural diffusion by trade may explain the western European transition equally well; (3) migrating farmers apparently contribute less than local adopters to the establishment of agropastoralism. Our study thus underlines the importance of adoption of introduced technologies and economies by resident foragers.
\end{abstract}

\end{frontmatter}

\section{Introduction}

The transition to agropastoralism in western Eurasia between 10\,000 and 3000\,cal\,BC was associated with enormous cultural, technological and sociopolitical changes.  Growing crops and herding animals have profoundly changed and continue to change global human history \citep[e.g.][]{Roth1887,Westropp1872,Diamond2002,Mithen2003,Barker2006,Ruddiman2006,Kaplan2010,Kutzbach2010}.  These changes may be viewed positively as a trajectory of progress in the way it had been seen by nineteenth and early twentieth century evolutionists \citep{Westropp1872,Childe1936}, or it may be seen as a road to perdition as it was, for instance, considered by J.~\citet{Diamond1997}: `` \ldots\ a catastrophe from which we have never recovered.  With agriculture came the gross social and sexual inequality.''  However the interpretation, the transition to agropastoralism, often termed Neolithisation, constitutes a major period of change in the history of humankind.

\subsection{Archaeology}
Neolithisation is believed to have begun during the early Holocene in the Fertile Crescent, a mountainous region between the Levantine coast and the Zagros ridge \citep{Flannery1973}. Archaeobotanical, archaeozoological and archaeogenetic work has  demonstrated that all food crops and animals---except the dog---have their origins in and around the Fertile Crescent as a single founder region.  The assemblage making up the Neolithic Package  includes wheat, barley, rye, lentils,  peas \citep{Willcox2005}, and cattle, sheep, goat and pigs \citep{Luikart2001,Edwards2007,Larson2007,Zeder2008}.

While tendencies towards sedentism and storage of wild plants may already be interpreted from the Natufian data \citep{Boyd2006}, intensive cultivation and domestication  of both plants and animals gradually began during the Younger Dryas and  only fully developed during the early Holocene \citep{Zeder2008,Willcox2009}.   European agropastoralism is allochthonous  and its most likely origins are in the Fertile Crescent, where farming and herding began---still in mixture with a broad spectrum of foraging practices---during the tenth millennium cal\,BC \citep{Flannery1973,Kuijt2002}.

The wider expansion of agropastoralism started around 8500\,cal\,BC, approximately 1000 years after the first appearance of domesticated cereals in the Levant.  The first clear evidence for colonist farmers was found on Cyprus \citep{Peltenburg2000,Colledge2004,Willcox2005};  the expansion ended after  4000\,cal\,BC, when Neolithic sites emerged on the British isles and throughout northern Europe \citep{Sheridan2007,Whittle2007gop}.  Details of the intermediate region specific accounts of transitions have been collected, for example, by \citet[]{Price2000}, \citet{Whittle2007gop}, and \citet{Gronenborn2010}, including the prominent sixth~millennium linear pottery cultures of central Europe  \citep[LBK, e.g.~][]{Luening1994} and the funnel beaker culture of the northern European plains \citep[TRB, after 4500\,cal\,BC,][]{Midgeley1992}. Not only did agropastoralism spread to the northwest from the Near East centres but also eastward as far as the Indus valley \citep{Fuller2006}. 

The question as to why agropastoral life style spread has been recently connected to environmental variations and conflict resolution: \citet{Dolukhanov1973}, \citet{Weninger2009} and \citet{Gronenborn2009cft} suggested an emergence and spread of farming as a result of climate induced crises periods, during which it may have become necessary for groups to fission, i.e.\ to move from one location to the other to escape conflicts.

Two contrasting concepts on the mechanism of the spread of farming across western Eurasia have existed side by side. One suggests the introduction of the new agropastoral technologies through movements of people---migrations of any form; the other suggests a technology shift through indigenous adaptations and inventions fostered by culture contact---information dispersal of any form.  \citet{Zvelebil1998} discriminates seven spreading modes, for example~elite exchange or leap-frog colonisation, as combinations or intermediate forms of the two opposite spread mechanisms.

The acculturation or cultural diffusion model corpus has, in the more recent past, been applied by a number of post-processual archaeologists---more typically for the British isles but also for the continent \citep{Hodder1990,Thomas1991}. It may go back to a critique by, for example, \citet{Zvelebil1988} of a migrationist model proposed by \citet{Renfrew1987} and later by others \citep[e.g.][]{Bellwood2005}.  In a way connected to these models are those where farming or animal husbandry were seen as regional developments within Europe. Such indigenist scenarios have been proposed for southern France, northwest Africa or Greece \citep[e.g.][]{Geddes1980,Courtin1974,Winiger1998,Theocharis1973}.

Opposing this position is the one of migration, where the new technology and cultigens arrive from Anatolia and the Fertile Crescent into Europe through migrating people. This migrationist position goes back to \citeyear{Childe1925}, when V.~G.~\citeauthor{Childe1925} noticed a gradient in the spatiotemporal distribution of ceramics from western Eurasia emanating from the Fertile Crescent northwest into Europe. In a later publication, \citet{Childe1942} suggested that population pressure in the source region was the driver of this outmigration.  His position was supported by \citet{Ammerman1971,Ammerman1973}, who formulated  the `wave of advance model', which was based on the concept of demic diffusion. Today, scholars from a number of disciplines favour migrationist models both for people and for cultigens \citep{Sokal1991, Richards2003nte,Pinhasi2005,Edwards2007,Bramanti2009,Balaresque2010,Haak2010}. Archaeology, particularly continental European archaeology, sees evidence for more complex scenarios of migrations and local acculturation notably in western central Europe and France \citep[e.g.][]{Jeunesse2000,Bentley2002,Gronenborn2007ccs}, but acknowledges that long-distance contacts across western Eurasia did exist during the mid-Holocene and should, at least partly, have been maintained by the migration of people.  It is yet unclear exactly when migrations began and what the relative importance of acculturation and movement of people was.  It may not be ruled out that the large-scale population replacements around the Neolithic began with the onset of the sixth millennium cal\,BC \citep{Gronenborn2007bmn,Gronenborn2011}.

\subsection{Mathematical Models}
The spatiotemporal structure of the advance of farming in Europe was first and very coarsely quantified by \citet{Edmonson1961}, who estimated the speed of the agropastoral transmission frontier at $1.9$\,\kma.  Later,  \citet{Clark1965} and \citet{Ammerman1971,Ammerman1973} based their analysis on radiocarbon dates (in three areas, at 53 sites, 103~sites, respectively) to calculate the velocity of the appearance of farming practice in Europe along a southeast-northwest gradient. All three studies found an approximately linear relationship between temporal and spatial distance of European Neolithic sites to four Near Eastern sites, with a slope of approximately~1\,\kma.    
\citet{Pinhasi2005} confirmed this finding on a more extended data set of 765 sites; they calculated  a spreading rate between $0.8$ and $1.3$\,\kma; even when shortest-path distances are considered (longer than great circle distances because of the detour necessary from the Near East to Anatolia and from central Europe to Iberia),  a similar rate ($0.6$--$1.1$\,\kma) is found. 
A data set of 477~sites, including boreal European sites, was used by \citet{Davison2006,Davison2007,Davison2009} who simulate for a noninteracting agropastoral subsistence style Neolithisation and its speed with a reaction-diffusion model.  They too arrive at a mean speed of 1\,\kma\  into Europe.

\citet{Ackland2007} simulated the spread of farming by including a `hitchhiking' advantageous trait in their reaction-diffusion model.  This new lifestyle addition could either reflect the immigration of farmers, or a dynamical conversion of foragers into farmers upon contact.  Their model predicts that Neolithic farmers outplaced the indigenous population up to a line approximately connecting today's Venice--Prague--Warsaw--Moscow.  To the north and west of this line, the so-called converts adopted the new lifestyle by cultural diffusion.

Available evidence, in part originating from isotopic and genetic studies, points to a discontinuous expansion sequence for western Eurasia \citep{Guilaine2001,Gronenborn2009cft, BocquetAppel2009,Schier2009} during which short dynamic phases of long distance rapid expansions were followed by periods of stand-still with local or regional colonisation. Discontinuities in the Neolithic advance, however,  have not been hindcasted by the aforementioned models. One possible reason is that in the frameworks  provided by \citet[e.g.][]{Ackland2007} or \citet{Davison2006,Davison2009} important aspects are missing. These may comprise more detailed descriptions of the resources needed and used by the people, the influence of the local biogeographic suitability for farming or herding, and of temporal variation in resource availability.  Their models do not simulate any endogenous cultural, technical or agrarian development. All these factors may in principle accelerate or slow down the process of Neolithisation and lead to a more complex spatiotemporal pattern than may be predicted by simple reaction-diffusion models. 

\begin{figure}
\includegraphics[width=\hsize]{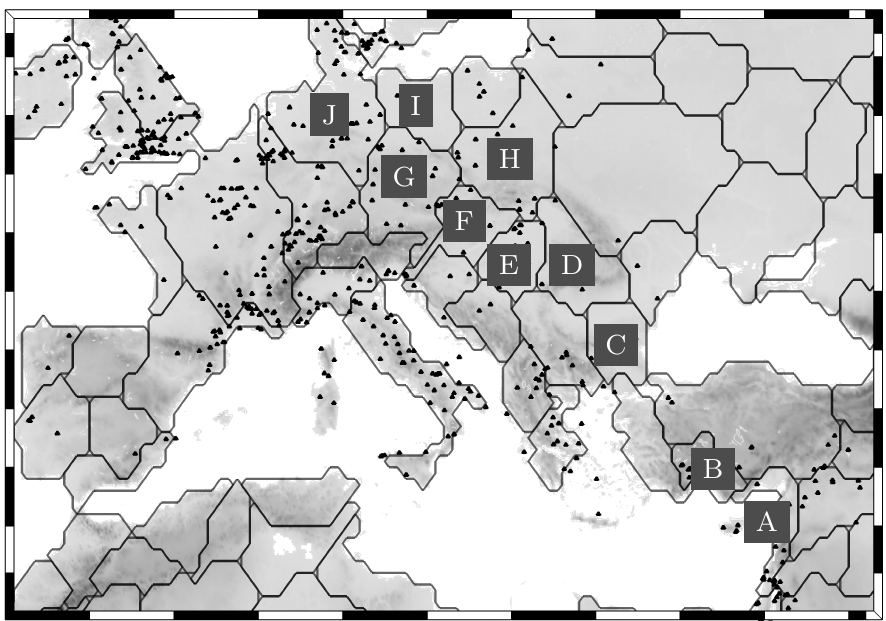}
\caption{Map of Europe with simulation region boundaries (solid lines)  in the Global Land Use and technological Evolution Simulator (GLUES).  Dots represent 631~radiocarbon dated Neolithic sites from \citet{Pinhasi2005}.  Our discussion focusses on a southeast to northwest transect along the highlighted regions A to J.}
\label{fig:regionmap}
\end{figure}

In this study, we employ the Global Land Use and technological Evolution Simulator (GLUES), which resolves local innovation, migration and cultural diffusion of traits \citep{Wirtz2003gdm,Lemmen2009,Lemmen2010}.  Although GLUES has been developed for the global domain, we restrict our analysis in this study to western Eurasia, where radiocarbon dates from Neolithic sites are abundant and of high quality, and where the issue of migration versus cultural diffusion is most intensively debated. The model is chosen because it allows us to differentiate between exchange (i.e.\ information exchange as cultural diffusion) and migration (demic diffusion) as important vectors of the expansion of agriculture. 

In the following section, we shortly introduce the GLUES model and the radiocarbon site data which are used for validation; a full description of the algorithms used in GLUES can be found in the supplementary online material (SOM).  The spatiotemporal pattern of the emergence and advancement of agropastoralism in western Eurasia is reconstructed and analysed in detail. This is achieved through a model-data comparison for ten focus regions along a southeast-northwest trajectory, from the Levant to north Germany; model-based expansion rates are put into the context of prior estimates from radiocarbon dates.  A major part of our discussion concentrates on the discrimination of migration versus trade and the maximum contribution of immigrants to emerging agropastoral communities in Europe. 

\section{Material and Methods}

GLUES mathematically resolves the dynamics of local human populations' density and characteristic sociocultural traits in the context of a changing biogeographical environment.   A local sociocultural coevolution is described by changes in mean population density, technology, share of agropastoral activities, and economic diversity, within a simulation region of approximately country-size extent (\reffig{fig:regionmap}).  Each local population utilises its regional natural resources, which are described by vegetation productivity and climatic constraints.  Each local population interacts with its geographical neighbours via trade and migration.  The conceptual model is outlined below, for details on the algorithms used and the mathematical implementation we refer the reader to \citet{Wirtz2003gdm}, summarised in the SOM.

\subsection{Characteristic traits}

For pre-industrial human societies, we define three characteristic traits:
\begin{enumerate} 
\item Technology is a trait which describes the efficiency of food procurement---related to both foraging and farming---and improvements in health care.  In particular, technology as a model describes the availability of tools, weapons, and transport or storage facilities. It aggregates over various relevant characteristics of early societies and also represents social aspects related to work organisation and knowledge management. It quantifies improved efficiency of subsistence, which is often connected to social and technological modifications  that run in parallel. An example is the technical and societal skill of writing as a means for cultural storage and administration, with the latter acting as a organisational lubricant for food procurement and its optimal allocation in space and among social groups.
\item  A second model variable represents the share of farming and herding activities, encompassing both animal husbandry and plant cultivation. It describes the allocation of energy, time, or manpower to agropastoralism with respect to the total food sector. 
We define a local population as Neolithic when this share is larger than the share of foragers---regardless of its technology, economic diversity, or population density.
\item Economic diversity resolves the number of different agropastoral economies available to a regional population.  This trait is closely tied to regional vegetation resources and climate constraints.  We do not, however, attribute specific plants and animals to each economy.  As an example, a value of four would be obtained when (1) domestic pigs and (2) goats and the growing of (3) barley and (4) wheat were present in a given population. A larger economic diversity offering different niches for agricultural or pastoral practices enhances the reliability of subsistence and the efficacy in exploiting heterogeneous landscapes.
\end{enumerate}

The temporal change of each of these characteristic traits follows the direction of increased benefit for success (i.e.\ growth) of its associated population; this concept had been derived for genetic traits in the works of \citet{Fisher1930}, and was recently more stringently formulated by Metz and colleagues \citep{Metz1992,Dieckmann1996dtc,Kisdi2010} as adaptive dynamics (AD).  In AD, the population averaged value of a trait changes at a rate which is proportional to the gradient of the fitness function evaluated at the mean trait value.  The AD approach was extended to functional traits of ecological communities \citep{Wirtz1996eve,Merico2009,Smith2011}, and was first applied to cultural traits of human communities by \citet{Lemmen2001} and \citet{Wirtz2003gdm}.
  
\subsection{Local resources}
Each simulation region is defined by a largely homogeneous vegetation productivity (measured as net primary productivity, NPP), resulting in an average size of \science{130}{3}\,km$^2$  (\reffig{fig:regionmap}). We reconstruct past distributions of NPP with a global climate model coupled to a vegetation module. 
\nobreak{Climber-2} \citep{Claussen1999} temperature and precipitation anomalies from the IIASA climatological data base \citep[International Institute for Applied Systems Analysis, ][]{Leemans1991} are converted to NPP according to the climate constraints on NPP from \citet{Lieth1975}; we do not use soil maps to constrain vegetation productivity.

From NPP, both the regional utility of natural food resources and the number of potential domesticates are derived. According to \citeauthor{Braidwood1949}'s (\citeyear{Braidwood1949,Braidwood1950}) hilly flanks hypothesis,  potential domesticates were most abundant in open woodlands at low to intermediate NPP{}. The number of potential domesticates furthermore depends on a continental aggregation to account for the area-biodiversity relationship \citep[e.g.][]{Begon1993}.  

\subsection{Exchange of information and people between regions}
Information exchange and migration are vectors of the spread of technology, economic diversity, and farming practice from the founding centres to adjacent simulation regions. We discriminate the diffusion of traits without involving resettlement of people  (cultural diffusion by  information exchange), and the diffusion of traits via migration (demic diffusion).  In GLUES, both mechanisms are driven by differences in influence between neighbouring local populations.

We assume that information travels two orders of magnitude faster than people. 
Exchange networks extend over  distances of up to 1000~km, in the later Mesolithic
and Neolithic \citep{Mauvilly2008,Gronenborn1999}; these networks were crossed many times during the active time---say ten years---of a Neolithic trader.   
Within this time span, a migration model like the one by \citet{Ammerman1973}, would allow for an advance of only ten km. This parameterisation leads to diffusivities for migration on the order of ten \sqkma, a value which is comparable to the diffusivities employed by other model studies of demic diffusion \citep[e.g.\ ][]{Davison2006,Ackland2007,Patterson2010}.  

\subsection{Reference data and simulated time scale}

Our reference data set is the comprehensive data collection of 765~sites by \citet{Pinhasi2005}.  These authors used site data provided by the United Kingdom Archaeology Data Service, the Central Anatolian Neolithic e-Workshop (CANeW), the radiocarbon CONTEXT database, and the Radiokarbondaten Online (RADON) database.  In their compilation, they included only sites with small dating uncertainty ($<200$\,a); they report dates as calibrated calendar years before present (relative to 1950) based on calibration of original $^{14}$C measurements with CalPal~2004.  This data set was created by the \citeauthor{Pinhasi2005}\ to provide a high quantity of dates and good spatial coverage at the expense of chronologic uncertainties, which could have been avoided, if for example only AMS-dated (accelerator mass spectroscopy) samples had been used; only few of the 765~sites, however, have been AMS dated.  For our purpose, this data set with many (possibly uncertain) dates represents the expansion of agropastoralism at a satisfactory level of detail.  Future simulations at a refined spatial scale would benefit from a data set with better chronologic control, where local and regional events are presented in higher resolution and where the regionally patchy nature of the expansion of agropastoralism is better represented.

From this data set, we choose for comparison those 631~sites which are located in the spatial model domain ($10^{\circ}$W--$42^{\circ}$E and  $31^{\circ}$N--$57^{\circ}$N) and the period of interest (8000--3500\,cal\,BC).  For each site, we use the age range computed from the reported calibrated radiocarbon age and the reported standard deviation.

For the mathematical model, we introduce the age scale `simulated time BC' (sim\,BC) to distinguish between empirically determined age models and the model time scale. Ideally, sim\,BC should be numerically equal to cal\,BC.

We set up the eight global model parameters such that  the simulation is able to hindcast an accurate timing and location of the early farming centres Fertile Crescent, northern China, and Mesoamerica \citep{Smith1997}, and a reasonable global pattern of the subsequent Neolithisation. 
The simulation is started at 9500\,sim\,BC. All of the 685 biogeographically defined regions (including 71 in western Eurasia) are initially set with farming activity at 4\% and established agropastoral communities at 0.25, what represents a low density Mesolithic technology population and a broad spectrum foraging lifestyle with low unintentional farming activity. The latter is assumed to represent early animal harvesting, selective seed gathering, and the active use of fire.

\section{Results}

\begin{figure}
\includegraphics[width=\hsize]{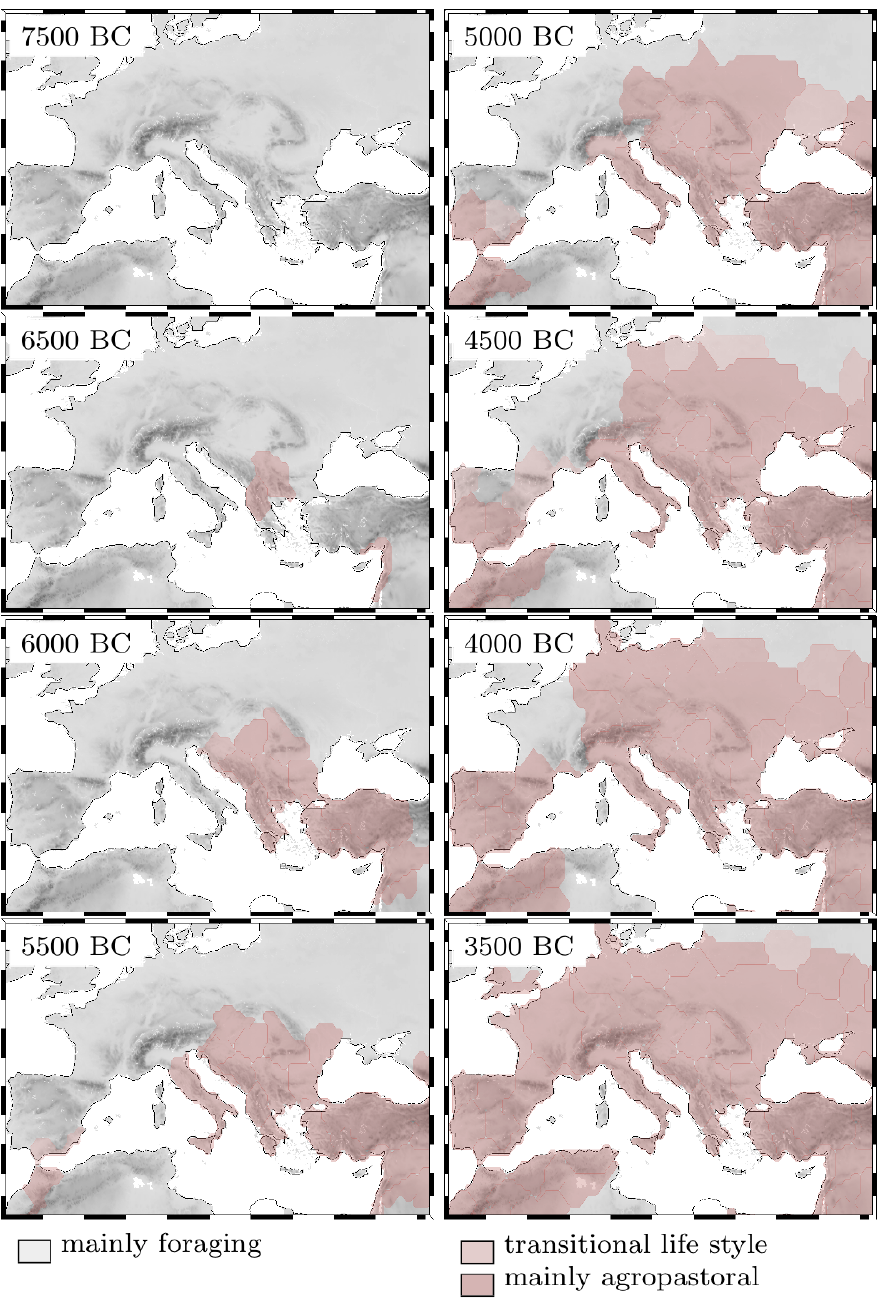}
\caption{The spread of agropastoralism in western Eurasia from 7500\,sim\,BC (top left, then downwards) until 3500\,sim\,BC, hindcasted with the Global Land Use and technological Evolution Simulator. A finer temporal resolution (50\,a time step) animation of this evolution is available in the SOM.}
\label{fig:farming}
\end{figure}

In the GLUES simulation, farming originates in the Levant (focus region~A, cmp.\,\reffig{fig:regionmap}) around 7000\,sim\,BC and penetrates into Europe in a northwest direction. By 3500\,sim\,BC all of continental Europe has converted to farming as the predominant subsistence style.  This emergence of farming in western Asia and Europe is shown as a series of snapshots in \reffig{fig:farming} (see SOM for an animated version with finer temporal resolution).  

\subsection{Expansion of agropastoralism}

The initial development progresses slowly and at a low level. It begins during the first century of the seventh millennium~sim\,BC in a region encompassing today's Lebanon, coastal Syria and a small part of the adjacent coastal Anatolia.  
In the 67th century~sim\,BC, northern Greece converts to agropastoral subsistence with rapid extension into the central Balkan. 
Over the next four hundred years, these agropastoral nuclei spread out further, encompassing the whole of Greece and the southern Balkan, and the coast of Anatolia by the 63rd century~sim\,BC.

A rapid expansion of agropastoralism occurs between 6200 and 6000\,sim\,BC, transforming the entire Balkan region and Anatolia. By 5750\,sim\,BC, the new subsistence mode has reached the northwestern and the easternmost coasts of the Black Sea.  In the 57th century sim\,BC, independent agropastoralism arises in north Africa in the region around the Strait of Gibraltar, and it emerges on the Italian peninsula.

The 55th century sim\,BC sees a rapid expansion of farming and herding into the area of the central LBK, and its spread into the south coast of the Iberian peninsula. By the 54th century, the LBK has expanded west- and eastwards and covers a vast stretch of land from southern Germany to the Ukraine; this central--eastern European area intensifies agropastoral activity without notable expansion until 5100\,sim\,BC.

Around 5000\,sim\,BC, forager societies on the north coast of the Black Sea, in north Africa and on the Iberian peninsula have converted to predominantly agropastoralism. At 4750\,sim\,BC, the Neolithic package reaches the Baltic Sea at the Oder river mouth; this coastal agropastoralism expands eastward until 4500\,sim\,BC and resembles the rise of the eastern TRB culture.  By this time, farming and herding have---in the model---reached the south coast of France and the north coast of Portugal.

An agropastoral area resembling the western TRB appears by 4400\,sim\,BC, also in southern Germany the new life style becomes dominant.  The later half of the fifth millennium sees a slow expansion towards the northeast of Europe, and the gap closure in central and northern France.  After 4000\,sim\,BC, agropastoralism reaches the British isles. 

\begin{figure*}
\includegraphics[width=\hsize]{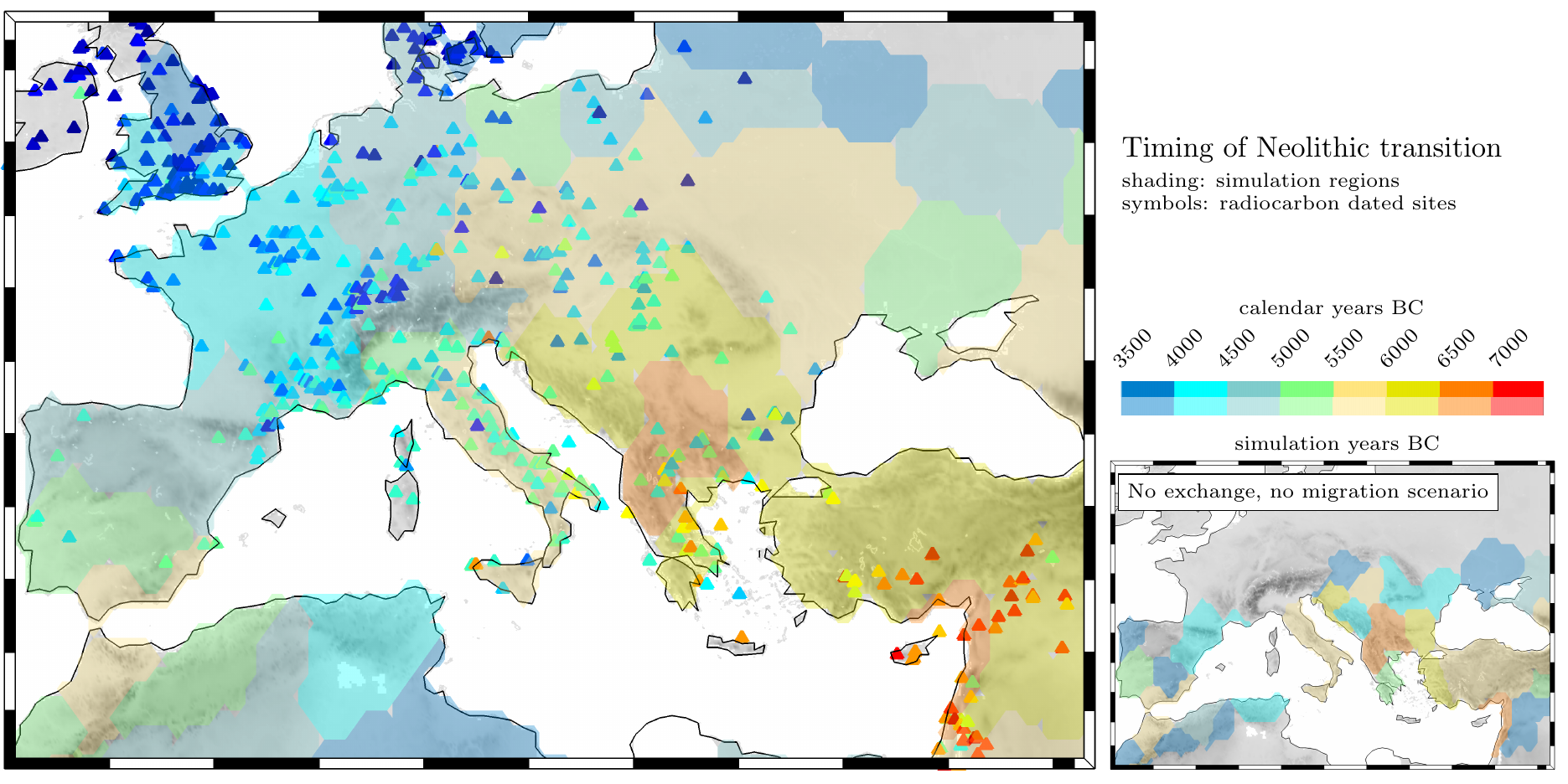}
\caption{Timing of the transition to agropastoralism in Western Eurasia.  The simulated transition (background pastel shading) is contrasted with the radiocarbon ages of Neolithic sites from  \citet[][solid colour triangles]{Pinhasi2005}.   The lower right inset image shows the transition for a scenario without migration or exchange, i.e, it shows the propensity of regions to endogenously develop agropastoralism.}
\label{fig:timingref}
\end{figure*}

\subsection{Timing of agropastoralism in model and data}
A summary description of the timing of agropastoralism between 7500 and 3500\,sim\,BC is  illustrated by \reffig{fig:timingref}; shown alongside are the median radiocarbon dates of Neolithic sites within this period from the data compilation by \citet{Pinhasi2005}.  From this time-integrated perspective, the simulated centres of agropastoralism in the Fertile Crescent, in northern Greece and at the Strait of Gibraltar are evident, as well as the southeast to northwest temporal gradient of the Neolithic transition.  The model-data comparison shows many good matches between radiocarbon dates and simulated transition dates.  We can clearly see, however, the spatial scale difference between simulation region and site data.  The spatial distribution of radiocarbon dated sites has good coverage along the transect from the Levant to northwestern Europe discussed below, it provides few or no information on eastern Europe, on the Iberian peninsula, and in north Africa.   

\begin{figure}
\includegraphics[width=0.95\hsize]{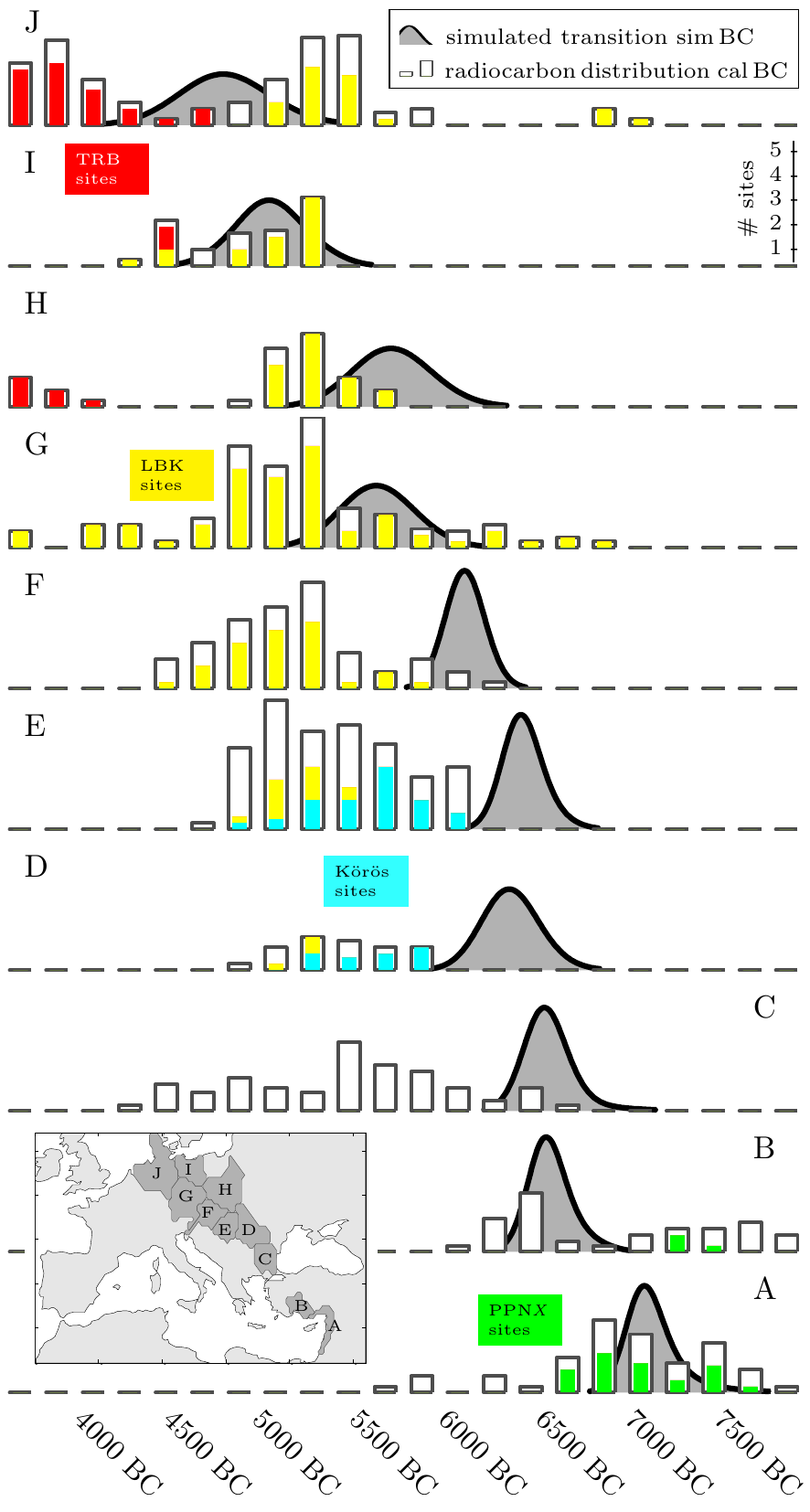}
\caption{Timing of the transition to farming in GLUES and calibrated radiocarbon date statistics for the ten focus regions (map inset).  In each panel A--I (ordered from southeast to northwest, bottom to top), the change in the fraction of agropastoralism from GLUES (grey shading) is contrasted to the number of radiocarbon dates compiled by \citet{Pinhasi2005} (bars, including 1$\sigma$ uncertainties of the age determination).  Colour indicates selected cultural attributions within the site record to pre-pottery Neolithic (of any kind, PPN\textit{X}), \koeroes, Linear pottery culture (LBK), and funnel beaker culture (TRB). To account for the scale difference, the simulated transition was broadened in time by a convolution with a normal distribution with mean zero and standard deviation proportional to region area and diffusion length. 
} 
\label{fig:timinghist}
\end{figure}

To assess the quality of the simulated onset of agropastoralism, we compare in \reffig{fig:timinghist} the change in fractional agropastoralism to the radiocarbon site statistics for ten focus regions A--H (a transect from the the Levant A to north Germany H) and radiocarbon dated sites within the region, or within 200\,km distance of the region centre for small regions.  We also indicate by colour selected cultural attributions.

Seventeen sites within or near region~A  are dated between 8000 and 5500\,\cal\,BC, of which the most frequent cultural attribution is Pre Pottery Neolithic (PPN$X$, 9~sites) and Pottery Neolithic (6~sites).  The most frequent century is the 68th\,cal\,BC (4~sites); the simulated change in agropastoral activity is greatest in the 70th century sim\,BC.

Ten sites are found in or near region~B, most of which are assigned to the Pottery Neolithic (6 sites).  All sites date to before 6000\,cal\,BC, with a maximum around 6400\,cal\,BC; the largest simulated change to farming occurs in this region and in region~C, around 6300\,sim\,BC.  Near or in C, 15~sites cover a wide temporal range from 6500 to 4400\,cal\,BC.  The site statistic within or around region~D is poor with only six sites, which date to 5800--5000\,cal\,BC. The timing of the largest simulated change is 6100\,sim\,BC;  this simulated transformation resembles the occurrence of the K\"or\"os culture. 

Like region~D, most sites near region~E are attributed to K\"or\"os; the second most frequent cultural complex in region~E is the LBK, which is also the dominant attribution at sites around regions F to~I.  In regions H to J, the site histogram is bimodal with the latter peak assigned to funnel beaker sites.  Many site dates near region~E fall within the period 6000--4800\,cal\,BC, whereas the simulated change is greatest at 6100\,sim\,BC;  around region~F, the simulated transition occurs around 5800\,sim\,BC. The most frequent date is the 52nd century cal\,BC, with a large range of 1500 years.  

Radiocarbon dates for region~G range from 6600 to 3500\,cal\,BC;  a maximum occurs between the 53rd and 47th century, which is concurrent with the largest simulated change at 5200\,sim\,BC.  Seven LBK sites around region~H are dated to 5600--5000\,cal\,BC, coterminous with the simulated shift at 5500\,sim\,BC.  For region~I, with mostly LBK-attributed sites and radiocarbon dates (5200--4400\,cal\,BC), the simulated subsistence change culminates around 4600\,sim\,BC.  The respective model transition for region~J appears at 4400\,sim\,BC. Here, the site histogram can be divided into two modes, where the first encompasses radiocarbon dates between  5800 and 4600\,cal\,BC (incl.~9 LBK sites) and the latter dates from 4200\,cal\,BC (incl.~10 TRB sites).

\subsection{Time lag--distance relationships}

\begin{figure}
\includegraphics[clip=,width=\hsize]{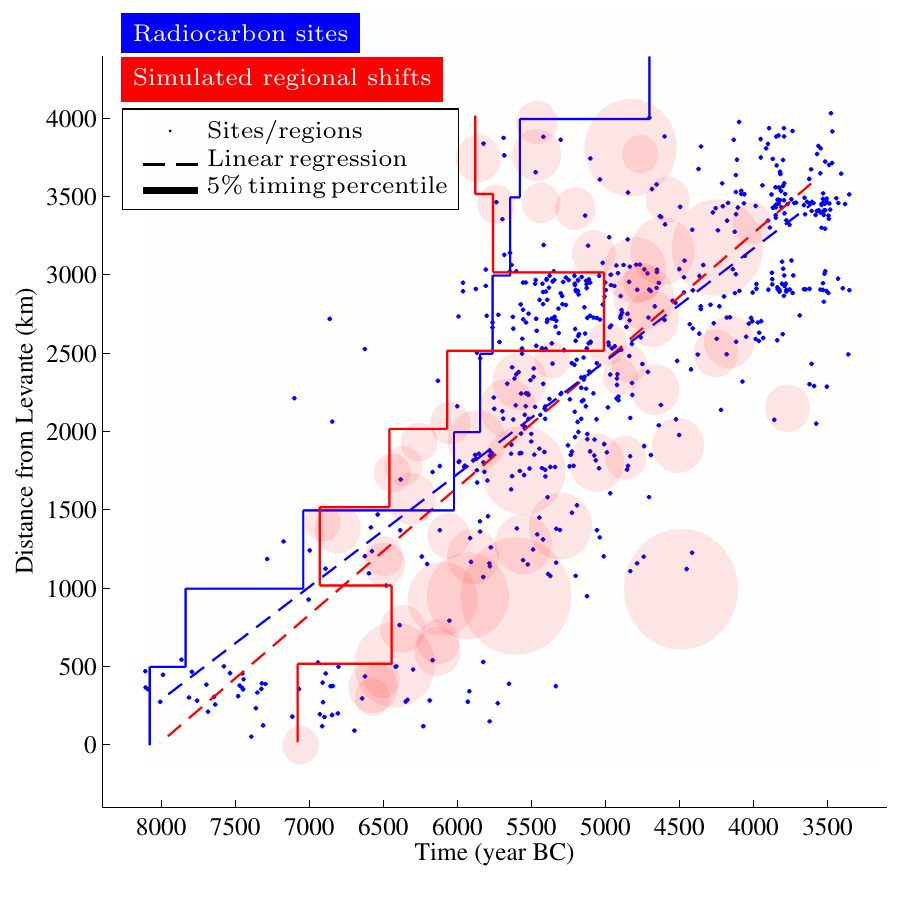}
\caption{Agricultural onset timing versus distance from an assumed archaeological centre near Beirut (Lebanon)  in the GLUES  simulation and in the data set by \citet{Pinhasi2005}.  
The correlation of distance and timing in the simulated data for $n=66$ regions is $r^2=.39$ (large circles, significant at $p=.01$), the average slope is $0.81$\,\kma.  For $n=631$ sites, the correlation is $r^2=.61$ (blue dots, significant at $p=.01$), the slope is $0.72$\,\kma.  The solid lines show for each distance interval of 500\,km the probability of the earliest and farthest occurrence ($p=0.05$ percentile) of agropastoralism in the site dates (blue) and the simulation (red).}
\label{fig:distance}­
\end{figure}

The average speed of the expansion of agropastoralism can be estimated from the time lag--distance relationship relative to an assumed founding centre of agropastoralism \citep[e.g.][]{Ammerman1973}.
\reffig{fig:distance} shows this relationship for all regions and radiocarbon dated sites within the model domain. Here, the assumed agropastoral centre is near today's Beirut, which lies in the middle of focus region~A. The (great circle) distances and time differences to this assumed centre extend over 4000\,km and 4000\,a, respectively.

Time lag and distance from the sites are highly correlated ($r^{2}=.61$) but also indicate a stair-case like distribution around a linear regression line with slope $0.72$\,\kma \citep[cmp.][]{Guilaine2001,Gronenborn2009cft,Schier2009}.  This means that in the data collection the spread of Neolithisation is slower in spatial proximity of the founder region than is predicted by a linear correlation; between 6000 and 4500\,cal\,BC, however, the graph of the majority of sites lies above the regression line, which indicates a more rapid wave of advance from the Balkan towards central Europe. Lag and distance for GLUES-simulated regions are also correlated to a marked degree ($r^{2}=.40$) and are similarly scattered around the regression line. Of the ten focus regions, regions~A and~B develop more slowly than expected from the regression and regions E--G develop agriculture faster than the linear regression. The average speed for the expansion of agropastoralism from the Levant into Europe calculated from the model is $0.81$\,\kma.

\section{Discussion}

The Global Land Use and technological Evolution Simulator is able to hindcast a realistic spatiotemporal pattern of the introduction of farming and herding into Europe between 8000 and 3500\,sim\,BC. The simulated expansion speed of agropastoralism compares well to a large dataset of radiocarbon dated Neolithic sites; the inhomogeneous spatial distribution of Neolithisation is reproduced. 

\subsection{spatiotemporal onset and expansion of agriculture}
The differences we observe between simulated timing and the radiocarbon age of sites within a simulation region (Figures~\ref{fig:timingref} and \ref{fig:timinghist}) are less than 1000\,a for almost all sites,  for the majority of sites less than 500\,a; only a handful of sites show differences greater than 1000\,a. These differences are similar in magnitude to those obtained by \citet{Davison2007} between their numerical model and radiocarbon dated sites in Europe. At this scale of model uncertainty, the radiocarbon dating uncertainty of individual sites ($<200$\,a) can be neglected:  the mismatch between the onset definitions in the data (presence of a Neolithic site, \reffig{fig:timinghist}) and in the model ($50\,\%$ agropastoral activity), as well as the spatial scale mismatch (local site data versus country-size simulation region, \reffig{fig:timingref}) introduce larger temporal differences.  To overcome the spatial scale problem, \citet{Zimmermann2004} argued for a landscape approach to archaeology, whereby a multitude of local sites are used to infer the archaeological context at the regional scale or larger.  The landscape approach can only succeed, however, if many sites within a region are excavated, as was the case for the lignite mining area of the Aldenhovener Platte studied by \citet{Luening1994} and \citet{Zimmermann2004}.  From a model perspective, more studies on methodologies to scale up the site (or many sites) information to the landscape are highly desirable. 
The scale difference illuminates the resolution limits of our model: GLUES resolves societal dynamics in larger environmental contexts rather than the history at individual sites. 

We find a marked correlation between the timing of first agropastoralism and the distance from a founding centre in the model ($r^{2}=.40$), and an average speed of agropastoralism in western Eurasia of $0.81$\,\kma.  Using radiocarbon data, a marked or high correlation was also found  by \citet[][$r^2=0.53, n=510$]{Gkiasta2003},  \citet[][$r^{2}=0.79, n=103$]{Ammerman1971}, and \citet[][$r^2=0.64, n=765$]{Pinhasi2005}. Differences between these empirical results can be attributed to the number of sites under consideration, to the location of the assumed founding centre, to site selection, or to the consideration of the shortest land route versus great circle distance \citep{Pinhasi2005}. For calibrated dates, \citeauthor{Pinhasi2005}\ calculate a speed range of $0.6$--$1.3$\,\kma{} when these differences are taken into account.  
The validity of comparing the onset of agropastoralism between simulated regions (with large areal extent) and (local) radiocarbon dated sites, despite the different scales, is supported by our simulation in two ways: (1) the marked correlation obtained between lag and distance of first agropastoralism to an assumed founding centre; (2) a calculated speed of $0.81$ that agrees closely with other published estimates.

Our simulations do not take coastal expansions into account, which would seem a major model deficiency at first glance.  The independent Moroccan model centre, however, acts for the Iberian peninsula similar as an explicit fast migration process \citep[like leapfrogging, on the order of $20$\,\kma,][]{Zilhao1993,Zilhao2000} along the Mediterranean coast and islands.  GLUES does not currently account for rivers:  we attribute the late transition of northern France in the simulation to a missing pathway from the Mediterranean coast through the Rhone valley. Indeed, \citet{Davison2006} found in their model a significant role of waterways in the Neolithisation of Europe,  whereas our results only indirectly (through the definition of regions by homogeneous vegetation) include river basins; GLUES performs well despite the lack of explicit river pathways like the Rhine or Danube valleys for all regions of Europe except central France. 

In the simulation, as well as in the data, the expansion of farming occurs in stages with periods of rapid spread followed by periods of local intensifications.  The rapid Neolithisation from Greece to the central Balkan in the 67th century sim\,BC is followed by a 400\,a period of relative stagnation. A very similar pattern is hindcasted for the  LBK-like Neolithisation in the 55th and 54th century cal\,BC, and for the relative stagnation before the onset of a TRB-like Neolithic further north.  Several regions exhibit a slow conversion to agropastoralism: according to \citeauthor{Bellwood2008}'s (2008) classification of zones in Neolithic Europe, France (see discussion of rivers above) would represent rather a friction than a spread zone. Another friction zone, where the Neolithic is introduced gradually, exists in the northern European lowlands, where empirical data supports the simulated late arrival of farming \citep{Midgeley1992,Zvelebil2006,Hartz2007}.

Our simulations predict a second and eastern expansion path around the Black Sea, which was archaeologically suggested by \citet{Kotova2003,Kotova2009}.  From a model perspective,  \citet{Davison2007,Davison2009} suggested that  to find a suitable simulation consistent  with the radiocarbon dated site context,  one needed an additional early wave of advance emanating around 8200\,cal\,BC from an Eastern European centre; agropastoral expansion would then proceed via the steppe corridor and be responsible for the eastern version of the Neolithic in Europe. This steppe corridor is archaeologically visible from the spread of pottery across Eurasia and the expansion of farming from eastern Anatolia into the Caucasus \citep{Dolukhanov2005,Gronenborn2009tcc,Kotova2003,Kotova2009}. 
In the reaction-diffusion migration model by \citet{Ackland2007}, a circular expansion from a single Mesopotamian centre is simulated; the geographical bottlenecks between the Mediterranean and the Black Sea, and between the Black and the Caspian sea act as secondary wave centres and thus the arrival of the farming wave in Europe ``appears to come from two sources: north and south of the Black Sea'' (\citeauthor{Ackland2007}, p.~8715). GLUES exhibits the same behaviour, with a southern and eastern path around the Black Sea based on a single Mesopotamian source area; for central Europe, however, our model suggests that the secondary centre is rather located in northern Greece and the central Balkan.  Archaeologically, the land route bottleneck may have not been too important for the Mediterranean coast, where the many islands provided a fast sea route from the Levant to Cyprus, Greece, and as far as Portugal \citep{Peltenburg2000,Theocharis1973,Arias1999ona,Zilhao2000}.

A separate non-Eurasian independent centre of agropastoralism is simulated by GLUES in the Maghreb.  From there, agropastoralism enters Europe via the Strait of Gibraltar around 5500\,sim\,BC \citep[cmp.][]{Manen2007}.  Archaeological records to show this are sparse; it is clear, however, that the strait had been in use as a migration path long since pre-Neolithic times, which is evident in gene pool analyses \citep[e.g.][]{Currat2010}.  Not only people but also domesticates crossed the strait in prehistory.
This was verified in a study by \citet{Anderung2005}, who found mitochondrial DNA of Bronze age (1800\,cal\,BC) Iberian cattle, of which a significant number possessed African haplotypes.

\subsection{Demic or cultural diffusion}

\begin{figure}
\includegraphics[width=\hsize]{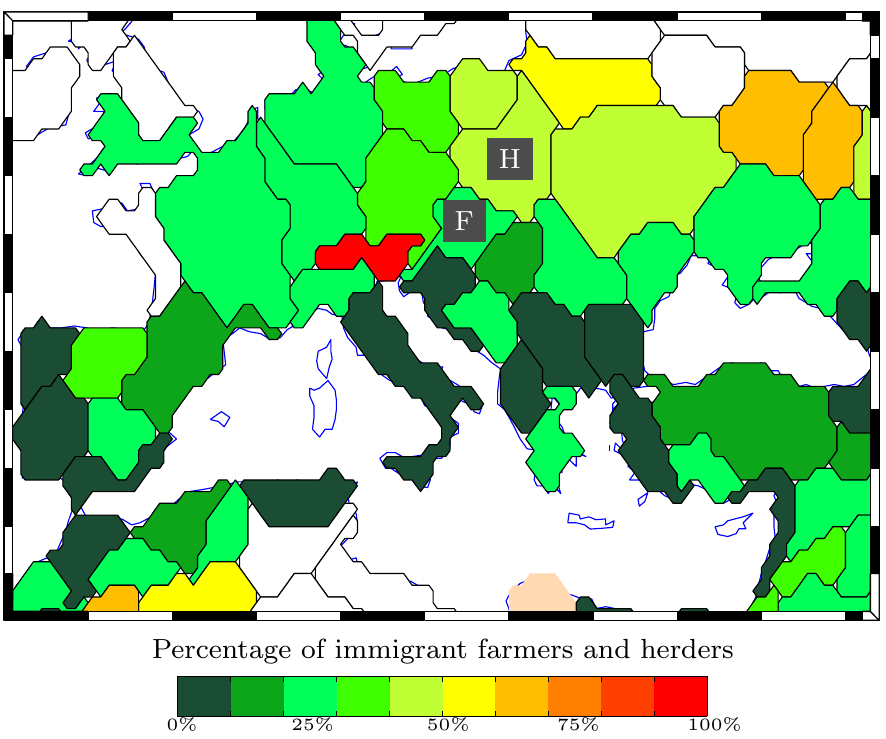}
\caption{Simulated fraction of immigrants in the agropastoralist population of each region at the time when the transition is locally 90\% complete.  Green colour indicates mainly local adoption by resident foragers, yellow to red a major contribution of immigrants. }
\label{fig:exchange}
\end{figure}

The relative contribution of demic versus diffusive processes can be calculated by following the streams of migration and trade in the model. We find that exchange processes contribute much less to local Neolithisation than adoption does.  In \reffig{fig:exchange}, we show this for the fraction of immigrant agropastoralists: for some (mostly Mediterranean) regions, immigrants are unimportant; for most regions, immigrants constitute one fourth of the agropastoral community; for a few northern and alpine regions, immigrants dominate.

We assessed the model sensitivity to different configurations for the speed of both information exchange and migration.  We find that (1) the model does not show sensitivity to either of these two parameters for a wide range of values, and that (2) the substitution fraction of agropastoralists with immigrants generally remained around 25\% for most regions; local invention and adoption of ideas dominate the Neolithisation, irrespective of whether people or information moved.

Is demic diffusion a \emph{sufficient} explanation for the European Neolithic? This has been suggested by many authors \citep{Ammerman1971,Ammerman1973,Sokal1991,Richards2003nte,Edwards2007,Bramanti2009,Balaresque2010,Haak2010}.  We can confirm this finding from a model simulation where cultural diffusion was deactivated.  Demic processes alone can reproduce the timing and lag-distance relationship seen in the radiocarbon data. 

But is demic diffusion \emph{necessary} for explaining the radiocarbon record?  \citet{Pinhasi2005} answered this question positively and pointed out that, at present, no working model existed that could explain the European Neolithic without demic diffusion.  Alike the demic diffusion-only experiment, we set up a simulation experiment where only information was allowed to diffuse, and where migration was inhibited: we could successfully reproduce the spatiotemporal emergence of the Neolithic in Europe with purely  cultural diffusion processes. To the question of necessity of demic diffusion for  the explanation of the radiocarbon record, our answer is no. This shows that the recently published evidence for major population transfers around the time of the Neolithisation process \citep{Haak2010} may have to be functionally disconnected from the spread of agropastoralism: apparently,  people did move at greater scales during the sixth and fifth millennium but these movements were not triggered by the spread of farming. It may entirely be possible that early---yet hypothetical---migrations already occurred before and during the seventh millennium and were undertaken by hunter-gatherers or mixed hunter-gatherer-horticulturalists originating from Anatolia. These immigrants then gradually pushed the original Mesolithic hunter-gatherer population of Europe towards the continental margins. Later, this migratory stream was complemented with farmers from Anatolia who then interacted with those hunter-gatherers who had arrived earlier \citep{Gronenborn2011}. This scenario would explain the archaeogenetic evidence for migrations as well as the archaeological evidence of interactions by disconnecting the spread of farming from the mid-Holocene migratory processes.

The insensitivity of the simulation results to the absence of either trade or migration processes prevents us from constraining the parameters for these processes quantitatively.  Even more, it tells us that from the phenology (the timing of agropastoralism) we cannot infer which of the two processes was responsible, or to which degree.  For the interpretation of radiocarbon dates of sites with an attribution to farming subsistence, one cannot find out whether demic or cultural diffusion was responsible for the apparent distribution in space and time of these sites. Or, put differently, the question on demic or cultural vectors may be not the most critical one for understanding the Neolithisation of Europe as a whole.

%
\begin{table}
\footnotesize
\begin{tabular}{l c c c }
& \parbox{7em}{\centering Region~F\\ agropastoralism\\ migr:adopt\\[1ex]}  
& \parbox{7em}{\centering Region~H\\ agropastoralism\\ migr:adopt\\[1ex]}  
& \parbox{7em}{\centering Region~H\\ technology\\ migr:exch:adopt\\[1ex]}
\\ \hline
demic    & 54:46 & 85:15 & 22:0:78 \\ 
mixed    & 22:78 & 41:59 & 6:13:79 \\
cultural  & 0:100 & 0:100 & 0:21:79 \\ \hline
\end{tabular}
\caption{Contributions (in percent) to local agropastoralism and technology from three different sources (1)  demic diffusion (labelled migr), (2) cultural diffusion (exch),  and (3) local adoption and invention (adopt) for three different model configurations with demic only, mixed (our standard configuration discussed in Figures~2--6), and cultural only diffusion.   Simulation results are shown for model regions corresponding to today's Hungary (focus region~F) and southern Poland (focus region~H).}
\label{tab:exchange}
\end{table}

What was the contribution of local adoption and invention? From \reffig{fig:exchange} it is evident that for most regions conversion of resident foragers to farmers played a larger role than immigration.  We quantitatively examined the relative importance of different sources (demic diffusion, cultural diffusion and local adoption or invention) to local Neolithisation in different model configurations (\reftab{tab:exchange}).  Even in a scenario where demic diffusion is the only active process, migration as a source does not explain 100\% of the agropastoralists in any focus region along the transect A--H but at most 85\% (in region~H),  less in regions closer to the Mediterranean coast (e.g.\ 54\% in region~F).  The local source (adoption and innovation) for an exchanged commodity like technology is in all configurations and focus regions more important (70--90\%) than migration or exchange.  

\subsection{Independent agropastoralism}
Was regional exchange (via migration or trade) necessary at all for the onset of agropastoralism everywhere in western Europe? In \reffig{fig:timingref} (lower right panel) we show the result of a simulation where both exchange processes were suppressed, thus endogenous transitions to agropastoralism become visible. The timing of the onset of agropastoralism around the Mediterranean Sea exhibits---next to the Levantine, Greek, and Moroccan centres which also appear in the reference simulation---many centres of hypothetical independent agropastoralism.  This independent agropastoralism is solely predicted on the basis of suitable environmental conditions (open vegetation type, not too cold) and internal development of sociocultural traits and demography;  it corresponds to indigenist scenarios proposed in the older literature, for example for southern France \citep{Geddes1980,Courtin1974} or Greece \citep{Theocharis1973,Winiger1998}. The indigenous agropastoralism hypothesis is, however, currently disregarded in archaeology because the genetic evidence points to the Near East as the centre for all Neolithic cultigens and nearly all domestic animals. 
For northern Europe, GLUES does not simulate the emergence of agropastoralism without the contribution of migration; these regions critically depended on the introduction of technologies and economies through the actual movement of people, commodities, and information.   

\subsection{Model comparison and outlook}
In addition to the prior approaches to simulating the European Neolithic  \citep{Davison2006,Davison2007,Davison2009,Ackland2007} which use geographic and topographic constraints for describing environmental heterogeneity, our model considers vegetation.  Vegetation production is directly coupled to the carrying capacity and  it determines the economic potential of a given environment.  Already many regions, mainly around the Mediterranean Sea, have a high propensity for developing independent agricultures based on the palaeoecological background (\reffig{fig:timingref}, lower right panel).  We couple the diffusion rates of traits and migration of people not only to the background geography but also to the (evolving) technology.  With these assumptions we can realistically reproduce the spatiotemporal pattern of Neolithisation in greater detail than was done for the front speed of Neolithisation by \citet{Davison2006,Davison2007} and \citet{Ackland2007}.

\citet{Ackland2007} and more recently \citet[][for the Indian transition to agropastoralism]{Patterson2010} use the concept of converts to describe resident foragers which have converted to agropastoralism.  We have shown that these converts may have played a larger role in the European Neolithisation than immigrant farmers.  Our model provides additional insight into the processes responsible for local adoption---often, a small share of introduced technology is sufficient to spark local invention and trigger the transition.  Alternatively, a few immigrant farmers and the technologies and economic possibilities they carry along may suffice to stimulate the local transition.  

Our regional prediction for western Eurasia emerges in the context of a global simulation: not only is the subcontinental prediction embedded in the larger spatial scale, but every local transition occurs within the temporal context of preceding predominant foraging subsistence with continuous innovation and succeeding intensification periods.  While we have shown that the model realistically reproduced the European Neolithisation, where archeological data is plenty and most reliable, the model's spatiotemporal consistency gives us confidence to draw conclusions about regions outside Europe in further studies. 

With the expected availability of more reliable palaeoclimate and palaeovegetation reconstructions from both models and data \citep{Kutzbach2010,Gaillard2010}, we expect to refine the large-scale biogeographic context of cultural evolution and the impact of local environmental disturbances \citep{Wirtz2010}.  We will then be able to assess better the degree to which the environment determined the potential transition to farming.  This potential should, however, be interpreted in G.~\citet{Ackland2007}'s way as providing a ``historical null hypothesis. Its predictions can be taken as requiring no special explanation, and its failures can be taken as evidence of rare events that had significant and long-lived consequences''. Numerical modelling of culture as a (natural) ecosystem may help to isolate the significant and non-deterministic events and concentrate our historical interpretation on those events where culture was most emancipated from the environment.

\section{Conclusion}

We presented a spatially explicit mathematical model of the Neolithisation of western Eurasia from 8000\,BC to 3500\,BC.  Our model incorporates endogenous sociotechnological dynamics, where culture is represented by the adaptation of characteristic population traits (technology, fraction of farmers, and economic diversity) and their interaction with demographics.  The study resolved the spatial expansion of Neolithic culture via indigenous development, migration and information exchange and reproduced the chronology of agropastoral onset observed in field data across western Eurasia, particularly reproducing and explaining the discontinuous speed of the `wave of advance'.

Our results encourage us to rethink possible indigenous centres along the northeastern shore of the Mediterranean: these might have not been able to develop since they were overrun by Near Eastern populations. Alternatively, the evidence for independent agropastoralism may have gotten lost in the admixture with the Fertile Crescent Neolithic package.  According to our simulations, a north African contribution to the European Neolithic should equally not be discounted.

The assessment of the relative importance cultural diffusion and demic diffusion in the model shows that either of these processes can explain the spatiotemporal pattern of agropastoral onset in Europe equally well. The phenology of the spatiotemporal pattern of agropastoral onset cannot discriminate the underlying process. Furthermore, even if only migration was considered and the diffusion of traits occurred only via immigrants,  the prevalence of immigrant farmers in any of the emerging agropastoral regions was much less than the prevalence of foragers who adopted the agropastoral life style. To the long-lasting dispute between cultural and demic diffusionists our novel interpretation offers a balanced explanation of predominant adoption despite migration. Whether the adopting population, however, did not also ultimately originate from Anatolia needs to be investigated by further archaeogenetic studies.

\section*{Acknowledgments}

We acknowledge the financial support for C.L.\ received from the Dutch Agency for Environmental Planning (Milieu- en Natuurplanbureau, De Bilt, The Netherlands, made possible by H. de Vries) and the German National Science Foundation (DFG priority project Interdynamik 1266).  We thank the two anonymous reviewers for their comments and J.~Jago for improving the language of the manuscript.

\bibliographystyle{elsart-harv}
\bibliography{journal-macros,glues}
\end{document}


\begin{frontmatter}

\title{Supplementary material for ``A simulation of the Neolithic transition in Western Eurasia''}

\author[hzg]{Carsten Lemmen\corref{lab:cor}}
\ead{carsten.lemmen@hzg.de}
\cortext[lab:cor]{Tel +49\,4152\,87-2013, Fax~-2020} 
\address[gkss]{Helmholtz-Zentrum Geesthacht, Institut f\"ur K\"ustenforschung, Max-Planck Stra\ss e~1, 21501~Geesthacht, Germany}
\author[rgzm]{Detlef Gronenborn}
\address[rgzm]{R\"omisch-Germanisches Zentralmuseum, Ernst-Ludwig Platz~2, 55116 Mainz, Germany}
\author[hzg]{Kai W. Wirtz}

\end{frontmatter}

\section{Supplementary movie}

\noindent Movie S1:  The spread of agropastoralism into Europe simulated with the Global Land Use and technological Evolution Simulator (GLUES) for the period 7500\,BC to 3500\,BC (cmp.\ main manuscript, Figure~2). The movie is encoded as MPEG-4, resolution 1428\,x\,1860 pixel, length 16~seconds (5~frames per second), file size 1.3~MB.

\section{Supplementary data}
The simulated fraction of agropastoral activity, the timing of the transition to agriculture, and the percentage of immigrant farmer data have been permanently archived on and are freely accessible from PANGAEA (Data Publisher for Earth \& Environmental Science, \href{http://www.pangaea.de}{www.pangaea.de}) as a netCDF dataset with reference  
``Lemmen, C et al. (2011): Simulated agricultural subsistence, timing of transition, and percentage of immigrant farmers.'' \href{http://doi.pangaea.de/10.1594/PANGAEA.761795}{doi:10.1594/PANGAEA.761795}.

\section{Supplementary information on GLUES}

The Global Land Use and technological Evolution Simulator \citep[GLUES, ][]{Wirtz2003gdm,Lemmen2009,Lemmen2010} mathematically resolves the dynamics of human populations and their characteristic sociocultural traits in the context of a changing biogeographical environment.   A local sociocultural coevolution is described by changes in mean population density (\density), technology (\T), share of agropastoral activities (\Q), and economic diversity (\N), within a simulation region of approximately country-size extent (Figure~1 of main manuscript).  Each region's population utilises its natural resources, which are described by environmental utility (\fep) and climate constraints (\tli).  While the mathematical implementation is summarised in its entirety from \citet{Wirtz2003gdm},  the reader is advised to refer to the original manuscript for detailed motivation for each model assumption.

\subsection{Characteristic trait dynamics}

For pre-industrial human societies, \citet{Wirtz2003gdm} defined three characteristic traits $X\in\{\T,\Q,\N\}$:
\begin{enumerate} 
\item Technology (\T) is a trait which describes the efficiency of food procurement related to both foraging and farming.
\item  A second model variable (\Q) describes the allocation of energy, time, or manpower to agropastoralism. 
\item Economic diversity (\N), resolves the number of different economies in the agropastoral sector which are available to a region's population.
\end{enumerate}

The evolution of each of these traits ($X$) follows the direction of increased benefit for success (i.e., growth benefit $\partial{\rgr}/\partial{X}$) of its associated population; this concept had been derived for genetic traits in the works of \citet{Fisher1930}, and more recently by Metz and colleagues \citep[e.g.][]{Kisdi2010} as adaptive dynamics (AD).  In AD, the population averaged value of a trait changes at a rate which is proportional to the gradient of the fitness function, evaluated the mean trait value:

\begin{equation}
\label{eq:xdyn}
\ddt{X} = \delta_X \, \del{\rgr}{X} \qquad\quad X\in\{\T,\Q,\N\},
\end{equation}
where $\delta_X$ is the so-called flexibility for trait $X$ and is often given by the variance
of $X$; 
 $\rgr$ denotes the relative growth rate of the population, i.e.\ $\rgr=(\mathrm{d}\density/\textrm{d}t)/\density$.

\subsection{Productivity and growth}

The Neolithic transition is characterised by changes in subsistence intensity (\si). Subsistence intensity describes a community's effectiveness
in generating consumable food and secondary products; this can be achieved based on an agricultural (with fractional activity \Q) and a hunting-gathering life style (with fractional activity $1-\Q$).
${\si}$ is dimensionless and scaled such that a value
of unity expresses the mean subsistence intensity of a hunter-gatherer
society equipped with tools typical for the mature Mesolithic and living in an affluent natural
environment.

\begin{equation}
\label{eq:food}
{\si}= (1- \Q)\cdot\sqrt{\T} +  \Q\cdot\N\cdot\T\cdot\tli
\end{equation}

The agricultural part of \si{} increases linearly with $\N$ and with $\T$:
The more economies (\N) there are, the better are sub-regional scaled niches utilised and the more
reliable returns are generated when annual weather conditions are variable; the higher the technology level (\T), the better the efficiency of using natural resources (by definition of $\T$).  While a variety of techniques can steeply increase harvests of
domesticated species, analogous benefits for foraging productivity are less
pronounced and justify a less than linear dependence of the
hunting-gathering calorie procurement on $\T$.  We use a square
root formulation, which satisfies $\sqrt{\T} <\T$
since $\T$ is generally larger than unity.

We introduce an additional temperature constraint (\tli) on agricultural productivity which considers that cold temperature could only moderately be overcome by Neolithic technologies. This limitation is unity at low latitudes and approaches zero at permafrost conditions.

The domestication process is represented by $\N$, which is the number of realised agropastoral economies. 
We link $\N$ to natural resources by expressing it as the fraction $f$ of potentially available
economies ($\pae$) by specifying $\N=f\cdot\pae$, where the latter  corresponds to the richness
in domesticable animal or plant species within a specific region.

\subsection{Food productivity and human growth rate}
\label{sec:growth}

The growth rate $\rgr$ of a regional population
relative to its size is mainly controlled by its subsistence intensity ${\si}$:

\begin{equation}
\label{eq:rgr}
r=\mu \cdot (\fep-\gamma\sqrt{\T}\density) \cdot (1-\omega\,\T)\cdot {\si}- \rho\cdot\density\cdot\mathrm{e}^{-\T/\T_{\mathrm{lit}}},
\end{equation}
with growth rate parameters $\mu$ and $\rho$.  The subsistence intensity's contribution to growth is modulated by environmental utility (\fep), societal impacts on the environment ($-\gamma\sqrt{\T}\density$), and organisational losses within a society ($1-\omega\,\T$).  As technology advances, more and more people neither farm nor hunt:  Construction, maintenance, administration draw a small fraction $\omega$ of the workforce away from food-production.   The impact on the environment is modelled as a function of population density and technology  \citep[IPAT,][]{Ehrlich1971} on the utility of the environment (\fep, see next section).
The loss term is mediated by technologies (\T, with $\T_\mathrm{lit}=12$), which mitigate, for example, losses due to disease.   The implementation of the loss term differs from the
originally inverse formulation by \citet[][$-\rho\frac{\density}{\T}$]{Wirtz2003gdm}, but is numerically simpler and phenologically similar.

\subsection{Region definition}
GLUES operates on a set of regions which, in turn, are defined on a high resolution ($0.5^\circ\times0.5^\circ$) grid. The region definition is implemented as a Cellular Automaton (CA), and resembles the application of a spatial low-pass filter to the grid. The CA formalism facilitates a dynamisation of region boundaries in future model applications. The CA aims at describing the organisation of local Mesolithic and Neolithic populations as cultural complexes that share common traits across larger spatial scales.
The algorithm combines single adjacent cells into a cluster (region), starting from a distribution where each cell forms an individual cluster. The CA iteratively computes the similarity in geographical characteristics between neighbour cells. Similarity is defined as the sum of absolute and normalised differences in discriminating local properties (\npp{} and growing degree days). The similarities to all neighbour cells are weighted by a factor $A/(A+A_T)$ where $A$ denotes the area of the adjacent cluster and $A_T$ a predefined target cluster size. The neighbour cell with maximal area weighted similarity then inherits cluster membership to the original cell. Cells that together with similar neighbours already form a cluster, usually do not change membership, while cells with a regionally anomalous biogeography become part of a new cluster. As a result, larger clusters incorporate cells at their boundary in relation to their own size and biogeographical similarity. The cluster distribution reaches a steady state when most cells are organised in clusters of relatively large cell number ($A>2A_T$).  Small clusters below the threshold size $A_T$ may, however, persist depending on the topography and the choice of $A_T$;  these are attached to one of the adjacent cluster regions using the area weighted similarity computed at the cluster scale.  

\subsection{Coupling to biogeography}
We reconstruct past distributions of net primary production (\npp) using a global climate model coupled to a vegetation module. From 
\nobreak{Climber-2} \citep{Claussen1999} temperature ($t$) and precipitation ($p$) anomalies on the IIASA climatological database \citep{Leemans1991}  we calculated \npp{} following
\citet[][Miami model]{Lieth1975}:

\begin{eqnarray}
&\npp_{p}   &=  \left(1-\mathrm{e}^{-0.664 p/\mathrm{m}}\right)\cdot\frac{1460\,\mathrm{g}}{\mathrm{m}^{2}\,\mathrm{a}} \nonumber\\
&\npp_{t}  &=  \left(1+ 3.7248\,\mathrm{e}^{-0.119 t / \mathrm{\degree C}}\right)^{-1}\cdot\frac{1460\,\mathrm{g}}{\mathrm{m}^{2}\,\mathrm{a}}\nonumber\\
&\npp  & = \min\left(\npp_{p},\npp_{t}\right)
\label{eq:lieth}
\end{eqnarray}

Local food extraction potential (\fep) represents the utility of  a multidimensional environment for food production; this \fep{} generally increases with \npp; at very high (i.e. tropical) \npp, it decreases because the amount of non-usable biomass rises \citep{Bloom1998} and the crop yield declines \citep{Gallup1999}.  We choose $\npp_{F}=1100\,\mathrm{g}\mathrm{m}^{-2}\,\mathrm{a}^{-1}$ to describe the productivity where \fep{} is highest and derive \fep{} as follows:

\begin{equation}
\label{eq:fep}
\fep = \dfrac{2 \frac{\npp}{\npp_{\mathrm{F}}}}{(\frac{\npp}{\npp_{\mathrm{F}}})^2 + 1}
\end{equation}

According to \citet{Braidwood1949}'s hilly flanks hypothesis,  potential domesticates were most abundant in open woodlands, which are characterised by low to moderate \npp; we accordingly use a monomodal transfer function to calculated the relative local potential for agropastoral economies (\lae), with a maximum at $\npp_{\N}=550\,\mathrm{g}\mathrm{m}^{-2}\,\mathrm{a}^{-1}$.

\begin{equation}
\label{eq:lae}
\lae = \tli\cdot\frac{4 \frac{\npp}{\npp_{\N}}}{(\frac{\npp}{\npp_{\N}})^3 + 3}
\end{equation}
Temperature limitation is included in the calculation of $\lae$ because cold-adapted plants
do generally not carry usable large seeds, fruits or tubers.

Following the concept of, for example, \citet[][]{Gaston2000} for the
connection of species richness on two different spatial scales, we connect the relative local potential ($\lae$) with the continental potential for 
agropastoral economies ($\cae$) to obtain an absolute local potential for agriculture ($\pae$); this way, we also account
for continental aggregation area-biodiversity relationships. Let $i$ denote a region index and $\mathcal{I}_{k}$ the set of regions on continent $k$.
Then

\begin{equation}
\pae_{i} = \lae_{i} \cdot \cae_{k} \qquad \mathrm{for} \qquad i\in\mathcal{I}_{k}
\end{equation}

As continents, we consider seven ($k\in\{1\ldots 7\}$) large biogeographic units: Australia, Subsaharan Africa, Eurasia including North Africa, North America, South America, Greenland, and one unit consisting of all large islands.
To obtain $\cae_{k}$, we calculate a weighted sum of all \lae{} with region area $A_{i}$ on continent $k$, and scale with the values for the largest continent (Eurasia)  where $\cae_{k}=\cae_{\max}$ and $A_{k}=A_{\max}$.  The consideration of region areas $A_i$ represents a linearised species-area relationship \citep{Begon1993}. 

\begin{equation}
\cae_{k}=\cae_{\max} \cdot A^{-1}_{\max}\cdot\sum_{i\in \mathcal{I}_{k}} \left(A_{i}\cdot\lae_{i}\right).
\end{equation}

\subsection{Interregional exchange}

Exchange between region $i$ and its neighbour $j$ in the neighbourhood $\mathcal{N}_{i}$ comprises communication, trade, colonisation, warfare, migration of individuals or entire populations.  We use the concept of influence by \citet{Renfrew1979edp}, defined by us as the product of $\density$ and \T{} to reflect imbalances in population pressure and disparate technologies.
Influence differences relative to the spatial average $\langle\T\cdot\density\rangle_{ij}$ act as a driving force $f_{ij}$ for exchange. 
The flux $f_{ij}$ is formulated as a first order relaxation \citep[e.g.][]{Wang1996} and depends on (1) the influence difference, (2) common boundary length $L_{ij}$ of the two regions, (3) the distance of their centres $\sqrt{A_{i}A_{j}}$, and (4) exchange coefficients for spread with people$\sigma_{\density}$ and without $\sigma_{\T}$. 

\begin{equation}
f_{ij}=\sigma_{\density,\T}\dfrac{L_{ij}}{\sqrt{A_{i}A_{j}}}\cdot\left(\langle\T\cdot\density\rangle_{ij}-\T_{i}\density_{i} \right)
\end{equation}

Whenever there is a positive flux $f_{ij}>0$, trait value differences in traits \T{} and \N{} from its neighbourhood $\mathcal{N}_{i}$ are added to region $i$'s traits:

\begin{equation}
\left.\displaystyle{\frac{\mathrm{d}X_{i}}{\mathrm{d}t}}\right\vert_{\T}
= \sum_{j\in\mathcal{N}_{i},f_{ij}>0} f_{ij}\cdot(X_{j}-X_{i})
\end{equation}

Population diffusion is mass-conserving and is composed of immigration ($f_{ij}>0$) and emigration ($f_{ij}<0$) of individuals or groups.  

\begin{equation}
\left.\displaystyle{\frac{\mathrm{d}\density_{i}}{\mathrm{d}t}}\right\vert_{\density}
= \sum_{j\in\mathcal{N}_{i},f_{ij}>0} f_{ij}\density_{j}\dfrac{A_{j}}{A_{i}}
- \sum_{j\in\mathcal{N}_{i},f_{ij}<0} f_{ij}\density_{i}
\end{equation}

With immigranting people, the characteristic traits of the migrants are carried over to the receiving region: 

\begin{equation}
\left.\displaystyle{\frac{\mathrm{d}X_{i}}{\mathrm{d}t}}\right\vert_{\density}
= \sum_{j\in\mathcal{N}_{i},f_{ij}>0} f_{ij}X_{j}\dfrac{\density_{j}A_{j}}{\density_{i}A_{i}}
\end{equation}

\bibliographystyle{elsart-harv}
\bibliography{journal-macros,glues}